\documentclass[twocolumn,fleqn]{edfest2005_elsart}

\usepackage{graphicx}
\usepackage{amsmath, amssymb, amsfonts}

\usepackage{natbib}

\newcommand {\version}{v3}
\newcommand{\versiondate}{October 2, 2006}
\renewcommand{\mathsf}{\text}         
\newcommand{\bR}{\mathbb{R}}

\newcommand{\dx}{\!\text{d}^4x\,\,}
\newcommand{\dint}[1]{\!\text{d}#1\,\,}

\newcommand{\dvy}{\!\text{d}^3 \vec y\,\,}
\newcommand{\tfr}[2]{{\textstyle \frac{#1}{#2}}}
\newcommand{\bdi}{\begin{displaymath}}
\newcommand{\edi}{\end{displaymath}}
\newcommand{\bfi}{\begin{figure}}
\newcommand{\efi}{\end{figure}}

\newcommand{\beq}{\begin{equation}}
\newcommand{\eeq}{\end{equation}}
\newcommand{\beqa}{\begin{eqnarray}}
\newcommand{\eeqa}{\end{eqnarray}}
\newcommand{\no}{\nonumber}

\newcommand{\sfd}{\text{d}}

\newcommand {\LC}  {Levi-Civita}

\newcommand {\rhs}    {right-hand side}

\newcommand{\lplanck}{l_\text{Planck}}
\newcommand{\Eplanck}{E_\text{Planck}}
\newcommand{\lgamma}{l_\gamma}

\newcommand{\lwormhole}{l_\text{wormhole}}
\newcommand{\lseparation}{l_\text{separation}}

\newcommand {\gsim}{\mathrel{\hbox{\rlap{\lower.55ex \hbox {$\sim$}}
            \kern-.3em \raise.4ex \hbox{$>$}}}}
\newcommand {\lsim}{\mathrel{\hbox{\rlap{\lower.55ex \hbox {$\sim$}}
            \kern-.3em \raise.4ex \hbox{$<$}}}}

\begin{document}
\begin{frontmatter}

\noindent
New. Astron. Rev. 54 (2010) 211
\hfill  astro-ph/0511267 (\version)\newline

\title{Spacetime foam and high-energy photons}
\author{Frans R. Klinkhamer}
\ead{frans.klinkhamer@kit.de\newline},
\author{Christian Rupp}
\ead{ch.rupp@gmx.de}

\address{Institute for Theoretical Physics,
     University of Karlsruhe (TH),\\
     76128 Karlsruhe, Germany}

\begin{abstract}
It is shown that high-energy astrophysics can provide
information on the small-scale structure of spacetime.
\end{abstract}
\date{\versiondate}
\end{frontmatter}

\mathindent=0mm
\section{Introduction}

The idea that spacetime on very small distance scales is
not perfectly smooth has a long history.
In modern times, the idea is often referred to as having a ``spacetime
foam'' instead of the smooth Minkowski
manifold \citep{W57,W68,H78,H91,V96,D04,Hu05}.

Several theoretical arguments for and against
a foam-like  structure of spacetime have been given.
(Spacetime would, for example, correspond to
a topologically trivial manifold, a multiply connected manifold,
a causal point set, a fermionic quantum vacuum, or something else.)
But,  ultimately, the question remains experimental:
\emph{what precisely \underline{is} the small-scale structure of spacetime?}

Needless to say, this  fundamental question is far from being answered.
However, it has been recently realized that high-energy astrophysics
may give valuable bounds or perhaps even clues.
See, e.g., \citet{A-C:dawn} and \citet{JLM:constraints} for two reviews.

The present contribution illustrates this astrophysics approach
by giving a brief summary of some of our own work
\citep{KlinkhamerRuppPRD,KlinkhamerRuppPRD-BR}. Of
course, this is a very subjective selection,
but it may, at least, give a concrete example
of some of the current research.

\begin{figure}[t]
\includegraphics[width=6.5cm]{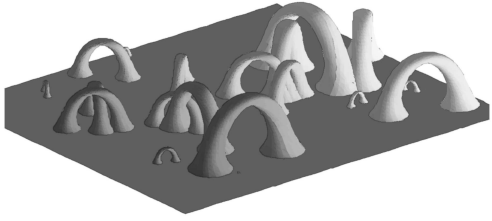}
\caption{View of a constant-time slice of a hypothetical version of
spacetime foam, with one spatial dimension suppressed. Illustrated is a
collection of ``wormholes'' \citep{W57}.  Points near the two
``mouths'' of an individual wormhole
can be connected either through the flat part of space
or though the wormhole ``throat'' (here shown as a tube rising above the
plane). The lengths of the wormhole throats can be arbitrarily small
(for a single wormhole, a visualization would require bending the plane).
\vspace*{0.125cm}}
\label{fig:foam}
\end{figure}

The procedure followed is in principle straightforward and
can be summarized by the following ``flow chart'':
\begin{description}
\item
\textsc{start:} assume a particular small-scale structure of spacetime
or, at least, one characteristic
(see Fig.~\ref{fig:foam} for an artist's impression);
\vspace*{0\baselineskip}
\item
\textsc{step 1:} calculate the effective photon model;
\vspace*{0\baselineskip}
\item
\textsc{step 2:} calculate the  modified photon dispersion
relation in the limit of large wavelengths;
\vspace*{0\baselineskip}
\item
\textsc{step 3:} compare with the astrophysical data and, if necessary,
return to \textsc{start}.
\end{description}
In this write-up, we only sketch \textsc{steps} 1 and 2 and focus on
\textsc{step} 3.
The crucial points are presented in the main text and
some back-of-the-envelope derivations
of the experimental limits are relegated to the appendices.

\section{Photon model and dispersion relation}
\label{sec:photon-model}

\textsc{Step} 1 mentioned in the Introduction
is conceptually and technically the most difficult of the
three. But this calculation is far from being completed
and can as well be skipped here. We simply
introduce a ``random'' (time-in\-de\-pen\-dent) background field $g$
to mimic the anomalous
effects of a multiply connected (static)
spacetime manifold with punctures, generalizing the result
for a single wormhole \citep{KlinkhamerRuppPRD}.
The physics of this type of anomaly has been reviewed in
\citet{K05}.

The photon model is now defined in terms of the standard gauge field
$A_\mu(x)$ over the \emph{auxiliary} manifold $\bR^4$ with Minkowski metric
$(\eta_{\mu\nu}) \equiv \text{diag}(1,-1,-1,-1)$.
(The real spacetime manifold
$\text{M}$ is assumed to look like Fig.~\ref{fig:foam},
but here we are only interested in long-distance effects
and approximate $\text{M}$ by $\bR^4$.)
The  Minkowski line element,
$\sfd s^2 =\eta_{\mu\nu}\: \sfd x^\mu\, \sfd x^\nu =
c^2\,\sfd t^2 - |\sfd\vec x|^2$,
defines implicitly the fundamental
velocity $c$ which need not be  equal to the (frequency-dependent) light velocity.

Specifically, the model action is given by
\begin{align}
&  S_\text{photon} =    \no\\[2mm]
&   -{\textstyle\frac{1}{4}} \int_{\bR^4}\: \dx\,
  \Bigl( \eta^{\kappa\mu}\,\eta^{\lambda\nu} \:
  F_{\mu\nu}(x)\,  F_{\kappa\lambda}(x) \no\\[2mm]
&  + \; g(x)\,  F_{\kappa\lambda}(x) \, \widetilde
  F^{\kappa\lambda}(x)\Bigr)\, ,
\label{Sphoton}
\end{align}
in terms of the Maxwell  tensor and its dual,
\begin{subequations}
\begin{align}
F_{\mu\nu}(x) &\equiv \partial_\mu A_\nu(x) -\partial_\nu A_\mu(x)\,,\\[2mm]
\widetilde F^{\kappa\lambda}(x) &\equiv
\tfr{1}{2}\;\epsilon^{\kappa\lambda\mu\nu}\, F_{\mu\nu}(x) \,,
\end{align}
\end{subequations}
with the completely antisymmetric
\LC~symbol $\epsilon^{\kappa\lambda\mu\nu}$ normalized by
$\epsilon^{0123}=1$.

With time-independent ``random'' couplings $g=g(\vec x)$
in the model action \eqref{Sphoton} and a few technical assumptions,
the modified dispersion relation is found to be given by:
\beqa
\omega^2 &=& (1-A^2 \, a_\gamma)  \: c^2 \, k^2
               -  \, A^2 \, l_\gamma^2 \:c^2\,k^4\no\\[2mm]
         &&   +\, \cdots \;,
\label{disp_foam}
\eeqa
for wave number $k\equiv |\vec k|\,$.

The constants $a_\gamma$ and $l_\gamma$ in \eqref{disp_foam} are
\emph{functionals} of the random couplings $g(\vec x)$:
\begin{subequations}\label{gamma1lfoam}
\begin{eqnarray}
a_\gamma &=&
\frac{\pi}{18\,A^2} \; C(0) \,,
\label{gamma1}
\\[2mm]
l_\gamma^2 &=&
\frac{2\pi}{15\,A^2}\; \int_0^\infty \dint{x} x\,C(x)\,,
\label{lfoam}
\end{eqnarray}
\end{subequations}
in terms of the isotropic autocorrelation function
\begin{subequations}
\beq
C(x)\equiv \widehat{C}(\vec x)\,,\;\; \text{for}\;\; x=|\vec x|   \,,
\eeq
with general definition
\begin{align}
\widehat{C}(\vec x) \equiv &\lim_{R\to \infty}\; \frac{1}{(4\pi/3) R^3}
\int\limits_{|\vec y|<R} \!\dvy
\no\\[2mm]
&\times g(\vec y) \, g(\vec y + \vec x)\,.
\label{correl}
\end{align}
\end{subequations}
For later use, we have also introduced an
``amplitude'' $A$ of the random couplings $g(\vec x)$,
with perhaps $A \sim\alpha$ from \textsc{step} 1. In this way,
the quantities $a_\gamma$ and $l_\gamma$
from (\ref{gamma1lfoam}ab) are independent of the overall
scale of $g(\vec x)$.

Note that the calculated dispersion relation \eqref{disp_foam}
does not contain a $k^3$ term, consistent with general
arguments \citep{Lehnert2003}.
The result \eqref{disp_foam} corresponds to \textsc{step} 2
mentioned in the Introduction.

In order to prepare for the first type of experimental limit, we
calculate the group velocity $v_g(k)\equiv \mathsf{d}\omega/\mathsf{d}k$
from (\ref{disp_foam}).
The relative change between wave numbers $k_1$ and $k_2$ is then
found to be given by
\begin{eqnarray}
\left.\frac{\Delta c}{c}\,\right|_{\, k_1,k_2} &\equiv&
  \left| \frac{v_g(k_1)-v_g(k_2)}{v_g(k_1)} \right| \no\\[2mm]
  &\sim& \, (3/2)
  \left| k_1^2 - k_2^2 \right| \, A^2 \, l_\gamma^2\,,
\label{theo}
\end{eqnarray}
where $\Delta c/c$ is a convenient short-hand notation.

\section{Experimental limits}
\label{sec:experimentallimits}

In this section, we obtain bounds from two ``gold-plated'' events,
a TeV gamma-ray flare from an active galactic nucleus (AGN)
and an ultra-high-energy cosmic ray (UHECR) from an
unknown source.

\subsection{TeV $\gamma$--ray flare}
\label{sec:Mkn421limit}

At the end of Section~\ref{sec:photon-model}, we have calculated the
relative change of the group velocity between two different
wave numbers $k_i$ (or photon energies
$E_i = \hbar\, \omega_i \sim \hbar\, c\, k_i$).
Following the suggestion of \citet{A-Cetal1998},
this theoretical result can be compared with astronomical
observations.

In fact, a particular TeV gamma-ray flare of the AGN Markarian 421
provides the following upper bound on $\Delta c/c\,$ \citep{Biller}:
\begin{equation}
\left.\frac{\Delta c}{c}\:
\right|^\mathsf{Mkn\,\, 421}_{
     \begin{array}{l}
     {\scriptstyle k_1=2.5\times 10^{16} \, \mathsf{cm}^{-1}} \\[-4mm]
     {\scriptstyle k_2=1.0\times 10^{17} \, \mathsf{cm}^{-1}}
     \end{array}}
\; < \, 2.5 \times 10^{-14} \,.
\label{exp}
\end{equation}

Combining the theoretical expression (\ref{theo}) and the astrophysical
bound (\ref{exp}) then gives the following ``experimental''
limit on the photonic length scale \citep{KlinkhamerRuppPRD}:
\begin{equation}
\lgamma < \left(1.8 \times 10^{-22}\;\text{cm}\right)\;\left(\alpha/A\right)  \;,
\label{lgammaboundMKN421}
\end{equation}
with fine-structure constant $\alpha \approx 1/137$ inserted for
amplitude $A$.

See App.~\ref{AppendixTime-dispersion} for some
details on this experimental limit and for a rough estimate of
what might ultimately be achieved with, e.g.,
the Gamma-ray Large Area Space Telescope (GLAST). For now,
bound \eqref{lgammaboundMKN421} is our first result for \textsc{step} 3
mentioned in the Introduction.

\subsection{UHECR}
\label{sec:UHECRlimit}

Recall the modified photon dispersion relation (\ref{disp_foam}) and,
for definiteness, assume an unmodified proton dispersion relation
$E_p^2=\hbar^2\,c^2\,k^2+m_p^2\,c^4$.

The absence of Cherenkov-like processes $p \to p \gamma$
\citep{ColemanGlashow1997} for a proton energy of the order of
$E_p  \approx 3 \times 10^{11} \, \text{GeV}\,$ then
gives the following experimental
limits \citep{GagnonMoore,KlinkhamerRuppPRD-BR}:
\begin{subequations}\label{gamma1lfoambounds}
\begin{align}
a_\gamma &< \left(6\times 10^{-19}\,\right)\left(\alpha/A\right)^2 ,
\label{gamma1bound} \\[2mm]
\lgamma &<  \left(1.0 \times 10^{-34}\,
  \text{cm}\right)\left(\alpha/A\right). \label{lfoambound}
\end{align}
\end{subequations}

See App.~\ref{AppendixCherenkov} for the basic
physics and astronomy input behind these limits.
Bounds (\ref{gamma1lfoambounds}ab) are the last results
for \textsc{step} 3 mentioned in the Introduction.

\subsection{Three remarks}

Having obtained these experimental limits, three remarks are in order.
First, bounds (\ref{gamma1bound}) and (\ref{lfoambound})
arise from soft ($\lesssim \text{GeV}$)
and hard ($10^{11}\,\,\text{GeV}$) photons, respectively,
whereas bound \eqref{lgammaboundMKN421}
comes from photons with intermediate energies ($10^{3}\,\,\text{GeV}$).

Second, bound  (\ref{lfoambound})
is twelve  orders of magnitude better than (\ref{lgammaboundMKN421}).
In App.~\ref{AppendixTime-dispersion}, we show that the
potential time-dispersion limit from GLAST
would still be far above the Cherenkov limit (\ref{lfoambound}).
This illustrates the power of using ultra-high-energy particles
\citep{ColemanGlashow1999}, at least for the present purpose.

Third, the Large Hadron Collider (LHC) at CERN will directly probe distances
of order $10^{-18}\, \text{cm}$,
far above the limits from astrophysics. But, then, there is nothing
better than a controlled experiment. Clearly, high-energy astrophysics
and experimental particle physics are complementary
in determining the small-scale structure of spacetime.

\vspace*{-0cm}
\section{The unbearable smoothness of space}

Purely mathematically, define
\begin{subequations}
\label{lfoamgamma1generaldef}
\begin{align}
\lgamma &\equiv \lwormhole\;
\left( \frac{\lwormhole}{\lseparation} \right)^{3/2},
\label{lfoamgeneraldef}\\[2mm]
a_\gamma &\equiv \left( \frac{\lwormhole}{\lseparation} \right)^3.
\label{gamma1generaldef}
\end{align}
\end{subequations}
For the moment, $\lwormhole$ and $\lseparation$ are just new symbols.

But, physically, the length $\lwormhole$ might correspond to
an appropriate characteristic dimension of an \emph{individual}
wormhole (e.g., the typical flat-space distance between
the centers of the mouths) and the length $\lseparation$  to
an average separation between \emph{different} wormholes
\citep{KlinkhamerRuppPRD}.
See Fig.~\ref{fig:foam}, but keep in mind that, most likely,
the real spacetime manifold cannot be viewed
as being embedded in a pseudo-Euclidean space.

\begin{figure}[t]
    \begin{center}
    \includegraphics[width=6cm]{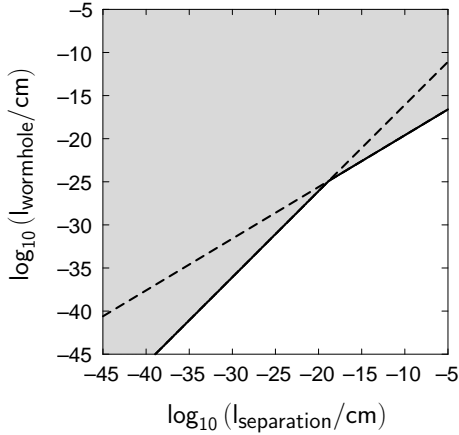}
    \end{center}
    \caption{Excluded  region [shaded] for
    random-coupling amplitude $A$ $=$ $\alpha$ $\approx$ $1/137$ (see text).
    \vspace*{0cm}}
\label{fig:lfoambounds}
\end{figure}

From bounds (\ref{gamma1bound},\ref{lfoambound})
and definitions (\ref{lfoamgamma1generaldef}ab), one
gets the exclusion plot of Fig.~\ref{fig:lfoambounds}.

It is perhaps not unreasonable to
expect  some remnant  ``quantum-gravity'' effect
with \emph{both} length scales $\lwormhole$ and $\lseparation$ of the order
of the fundamental Planck length \citep{W68},
\beqa
\lplanck &\equiv&  \hbar\, c/\Eplanck \equiv \sqrt{G\,\hbar/c^3} \no\\[0mm]
         &\approx& 1.6 \times 10^{-33}\,\,\text{cm}\,,
\eeqa
with $\Eplanck \approx 1.2 \times 10^{19}\,\,\text{GeV}$.
However, $\lwormhole \sim \lseparation \sim 10^{-33}\,\,\text{cm}$
seems to be  ruled out by the limits shown
in Fig.~\ref{fig:lfoambounds},
provided the amplitude $A$ of the effective random coupling
$g(x)$ in model \eqref{Sphoton}
is indeed of order $\alpha\approx 1/137$, as suggested
by preliminary calculations \citep{KlinkhamerRuppPRD}.

The \emph{tentative} conclusion is that a preferred-frame graininess of
space with a single length scale $\lplanck$ may be hard to reconcile
with the current experimental bounds from cosmic-ray physics.

If this conclusion is born out, the question becomes:
\vspace*{0.\baselineskip}\newline
\emph{Why is ``empty space'' so remarkably smooth?}
\vspace*{0.\baselineskip}\newline
An alternative form of the same question might read:
\vspace*{0.\baselineskip}\newline
\emph{Why is Lorentz invariance such an accurate symmetry of Nature,
      perhaps even up to energies of order $\Eplanck$?}

For further discussion on the question in the second form,
see, e.g., \citet{CorichiSudarsky} and \citet{KlinkhamerVolovik}.

Returning to the question in its original form
(which may also be related to the cosmological constant problem),
a direct experimental
solution would appear to be prohibitively difficult at present,
in view of the small numbers already appearing
in \eqref{lgammaboundMKN421} and \eqref{lfoambound}.
It may very well be that the only methods available are ``indirect,''
 either in the laboratory (e.g., neutrino oscillations)
or via astrophysics (e.g., GRBs and UHECRs).

\section*{Note added in proof}

The work reported on in this review article has been
continued over the last five years, with two main results.
\par
Following \textsc{steps} 1 and 2 mentioned in the Introduction,
the effects of a Swiss-cheese-type small-scale-structure
of spacetime on the pure photon theory have been calculated
in the long-wavelength limit.
The focus of this calculation was on the modification of the
photon propagation because of the presence of finite-size
holes or defects in the underlying spacetime manifold.
Without ``conspiracy'' of the spacetime holes/defects, the modified
photon  propagation is nonbirefringent (i.e., with equal phase
velocities of the two polarization modes).
See \citet{BernadotteKlinkhamer}.
\par
Following \textsc{step} 3 of the Introduction, tight
bounds on the nonbirefringent Lorentz-violating
parameters of modified Maxwell theory have been obtained
using data from \mbox{UHECRs} (Pierre Auger Observatory)
and TeV gamma-rays (\mbox{HESS} imaging atmospheric Cherenkov telescopes).
See \citet{KlinkhamerRisse,KlinkhamerSchreck}.

\ack
FRK thanks Robert D. Preece for a brief but informative
discussion on GLAST, the organizers for bringing about this
remarkable meeting, and, last but not least, Ed van den Heuvel
for his interest and support over the years.

\begin{appendix}
\section{Time-dispersion limit}
\label{AppendixTime-dispersion}

The basic idea \citep{A-Cetal1998} is to look for time-dispersion effects
in a burst-like signal from a very distant astronomical
source. The assumption is that the original event was really
sharply peaked in time for \emph{all} photon energies simultaneously,
excluding the hypothesis of some ``cosmic conspiracy.''

\begin{figure}[t]
\begin{center}
\includegraphics[width=6.5cm]{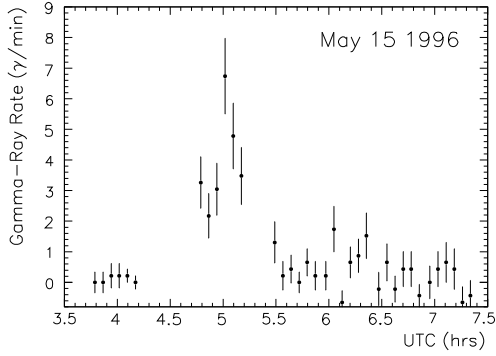}
\end{center}
\vspace*{.0\baselineskip}
\caption[Flare from Markarian 421]
{TeV $\gamma$--ray flare from Markarian 421 observed on
May 15, 1996, by the Whipple $\gamma$--ray observatory.
The rate of excess $\gamma$--ray selected events is binned
in intervals of 280 seconds. From \citet{Biller}.\vspace*{0.125cm}}
\label{may_flare}
\end{figure}

\begin{figure}[t]
\begin{center}
\includegraphics[width=6cm]{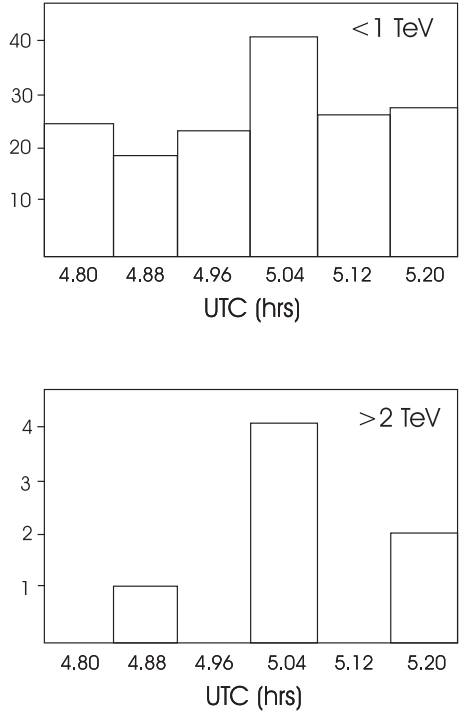}
\end{center}
\vspace*{.0\baselineskip}
\caption[Flare from Markarian 421]
{Total number of $\gamma$--ray selected events occurring in
each 280 second interval near the peak of the 15 May 1996 flare
from Markarian 421. The top plot consists of events with
$\gamma$--ray energies less than 1 TeV, whereas the bottom plot
is for energies greater than 2 TeV. From \citet{Biller}.\vspace*{0.125cm}}
\label{flare_comp}
\end{figure}

One particular event turns out to be most useful for our purpose and
the observations are shown in Figs.~\ref{may_flare}  and
\ref{flare_comp}. The three peaks of these two figures occur in the
same time bin and no time-dispersion is seen. The ratio of the
binning interval ($\Delta t \approx 280\: \text{s}\,$) over the
inferred travel time ($D/c \approx 1.1\times 10^{16}\:\text{s}\,$)
then gives bound (\ref{exp}) which, in turn, gives bound
\eqref{lgammaboundMKN421} on the photonic length scale $\lgamma$.

It may be of interest to see what time-dispersion limits can be reached
in the future. In order to be specific, we take
the most optimistic values for
GLAST \citep{McEnery-etal2004,Bhat-etal2004}.
This forthcoming satellite experiment
has an energy range up to several hundred GeV, may detect
gamma-ray bursts (GRBs) up to distances $D \sim 10^{10}\,\text{lyr}$,
and has a time resolution of the order of a few microseconds.

From \eqref{theo}, an upper bound $\Delta c/c = c\,\Delta t/D$
for photon energies $E_1 \ll E_2 \equiv E_\text{max}$ then
implies the following bound on the photonic length scale:
\mathindent=0mm
\beqa
&&\lgamma \leq
\frac{1}{\sqrt{3/2}\;A}\;
\left(\frac{\hbar\, c}{E_\text{max}}\right) \;
\left( \frac{c\,\Delta t}{D} \right)^{1/2}
\no\\[2mm]
&& \approx
1.9 \times 10^{-26} \:\text{cm}
\left( \frac{\alpha}{A} \right)
\left( \frac{300 \;\text{GeV}}{E_\text{max}} \right)
\no\\[2mm]
&& \times
\left( \frac{\Delta t}{2 \: 10^{-6}\;\text{s}} \right)^{1/2}
\left(  \frac{3 \:10^{17}\;\text{s}}{D/c} \right)^{1/2} \!\!\!.
\label{lgammaboundGLAST}
\eeqa
This estimate shows that GLAST may indeed be a remarkable probe of
small-distance physics (compared to $10^{-18}\, \text{cm}$ from the
LHC, for example). Note that
the effective length scale probed, $\lgamma$,
appears quadratically in the
dispersion relation \eqref{disp_foam}, which
explains the presence of square roots on the \rhs~of
\eqref{lgammaboundGLAST}.

\section{Cherenkov limits}
\label{AppendixCherenkov}

If the photon dispersion relation is modified, otherwise forbidden photon-radiation
processes may become kinematically allowed. We consider,
in turn, two different modifications of the photon dispersion relation.
Furthermore, we set $\hbar=c=1$ in this appendix and,
for simplicity, assume an unchanged proton dispersion relation,
$E_p^2=k^2+m_p^2$, with momentum $k \equiv |\vec k|$.

The first modification considered corresponds to a possible reduction of the
speed of light compared to the maximum attainable speed $c=1$ of the proton:
\begin{equation}
E_\gamma= (1-\epsilon)\, k\,,\quad  0\leq \epsilon <1 \,.
\label{case1}
\end{equation}
A charged particle traveling faster than light can now emit
``vacuum Cherenkov radiation'' \citep{ColemanGlashow1997}.

It is, of course, known that hadrons with particularly high energies
occur in cosmic rays. As vacuum Cherenkov radiation
would slow down charged  hadrons traveling through empty space,
no  hadronic cosmic rays with
velocities substantially above the speed of light
would ever reach the Earth's atmosphere.

\begin{figure}[t]
\begin{center}
\includegraphics[width=6.5cm]{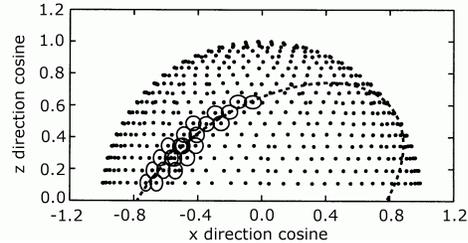}
\end{center}
\vspace*{0mm}
\caption{Pointing directions of the 22
photomultiplier tubes which triggered in connection
with the Fly's Eye event of October 15, 1991, at 7:34:16 UT.
The pointing directions are shown projected into the $x$--$z$ plane, where
the $x$--axis points east, the $y$--axis north, and the $z$--axis upward.
The triggered phototubes have positive $y$--components.
The dashed line indicates the plane defined by the shower axis
and the detector. From \citet{Bird-etal1995}.\vspace*{0.125cm}}
\label{UHECRflyseye}
\end{figure}

The most energetic cosmic ray reported so far was observed on
October 15, 1991,  by the Fly's Eye Air Shower Detector in Utah
and had $k\approx 3 \times 10^{11} \,\text{GeV}$
\citep{Bird-etal1995}. See Fig.~\ref{UHECRflyseye} for the pattern
of triggered photomultiplier tubes.
Note that several other events with $k \sim 10^{11} \,\text{GeV}$
have been observed by the Akeno Giant Air Shower Array
\citep{Takeda-etal1998}.

This particular Fly's Eye event \citep{Bird-etal1995,Risse-etal2004}
corresponds to a velocity of the assumed primary proton:
\begin{align}
v &\sim    1-  \textstyle{\frac{1}{2}} \, m_p^2 \, /\, k^2\no\\[2mm]
  &\approx 1- 6\times 10^{-24}\,,
\label{protonvelocity}
\end{align}
where $m_p \approx 1\, \text{GeV}$ is the proton mass.
By direct comparison of (\ref{protonvelocity}) and (\ref{case1}),
a rough upper bound on $\epsilon$ can be obtained. More precisely,
the partonic content of the proton has to be taken into account
\citep{GagnonMoore}, which leads to a somewhat weaker bound,
\begin{equation}
0 \leq \epsilon < 1.6 \times 10^{-23}\,.
\label{epsilon_bound}
\end{equation}
Identifying $\epsilon \equiv (1/2)\,A^2\,a_\gamma\,$,
bound \eqref{epsilon_bound}
gives  \eqref{gamma1bound} from the main text.

\begin{figure}[t]
\begin{center}
\includegraphics[width=4cm]{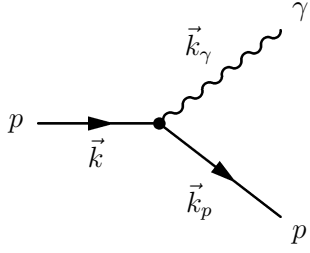}
\end{center}
\caption{Feynman diagram for proton vacuum Cherenkov radiation.
The heavy dot stands for a momentum-dependent form factor,
as the proton $p$ is a composite particle.\vspace*{0.125cm}}
\label{feynman}
\end{figure}

As the second modification of the photon dispersion relation, consider a
contribution to the photon energy which is cubic in the  momentum,
\begin{equation}
E_\gamma = k - K_{1}\, k^3\,.
\end{equation}
The Cherenkov-like process $p \, \rightarrow \, p\,\gamma$ (see
Fig.~\ref{feynman}) is then kinematically allowed if the proton energy
exceeds a particular threshold energy.

Given the direction of the initial proton momentum, one
defines the energy $E^\text{coll}$ of a decay particle to
correspond to the energy of the collinear part of its momentum.
Assume, first, that the three-momenta of the two
final-state particles are collinear (i.e., $E^\text{coll}=E$) and
that the photon carries away a fraction $x$ of the initial proton momentum. Due to
energy conservation, the difference between final and initial state energy has to
vanish, so that
\begin{equation}
\left(E^\text{coll}_\text{final} - E^\text{coll}_\text{initial}\right)
\Big|^{\begin{array}{l} {\scriptstyle \,\text{collinear}}\\[-5mm]
                        {\scriptstyle \,\text{case}} \end{array}}  =0 .
\end{equation}

For a finite opening angle of the final state particles, some energy is
used up by the transverse momentum components. One then has the inequality
\begin{equation}
\left(E^\text{coll}_\text{final} - E^\text{coll}_\text{initial}\right)
\Big|^{\begin{array}{l} {\scriptstyle \,\text{general}}\\[-5mm]
                        {\scriptstyle \,\text{case}} \end{array}}  \leq 0 .
\end{equation}
Taking $x=1/2$ and a value $k \approx 3 \times 10^{11}\,\text{GeV}$,
the process shown in Fig.~\ref{feynman} is allowed if
\beqa
 K_{1} &\ge&       5.7\,\, m_p^2 \, / \, k^4 \no\\[2mm]
       &\approx&  (4 \times 10^{22}\, \text{GeV})^{-2}\,,
\label{K1-threshold}
\eeqa
where, again, the partonic
content of the proton has  been taken into account \citep{GagnonMoore}.
Identifying $K_1 \equiv (1/2)\,A^2\,\lgamma^2\,$,
bound  \eqref{K1-threshold} gives \eqref{lfoambound} from the main text.

Remark, finally, that the bounds of this appendix and
Section \ref{sec:UHECRlimit}
were based on the assumption of having a primary proton $p$ (mass $m_p$)
for the Fly's Eye event considered
(see also recent results reported by \citet{Abbasi-etal2005}).
With a primary nucleus $N$ (mass $m_N$),
bounds \eqref{gamma1bound} and \eqref{lfoambound}
would increase by  approximate factors
$(m_N/m_p)^2$ and  $(m_N/m_p)$, respectively,
as follows by consideration of Eqs.~\eqref{protonvelocity}
and \eqref{K1-threshold}, neglecting form factors.
\end{appendix}



\vspace*{1.75cm}
\begin{thebibliography}{}


\bibitem[Abbasi et al.(2005)]{Abbasi-etal2005}
Abbasi, R.U., {\it et al.}  [The High Resolution Fly's Eye Collaboration],
2005,
Astrophys. J.  {622}, 910  [arXiv:astro-ph/0407622].


\bibitem[Amelino-Camelia et al.(1998)]{A-Cetal1998}
Amelino-Camelia, G., Ellis, J.R., Mavromatos, N.E., Nanopoulos, D.V., Sarkar, S.,
1998,
Nature {393}, 763  [arXiv:astro-ph/9712103].

\bibitem[Amelino-Camelia(2000)]{A-C:dawn}
  Amelino-Camelia, G., 2000,
  Lect. Notes Phys. {541}, 1 [arXiv:gr-qc/9910089].

\bibitem[Bernadotte and Klinkhamer(2008)]{BernadotteKlinkhamer}
Bernadotte, S., Klinkhamer, F.R.,
2007,
Phys. Rev.  D  {75}, 024028 [arXiv:hep-ph/0610216].

\bibitem[Bhat et al.(2004)]{Bhat-etal2004}
  Bhat, P.N., {\it et al.},
  2004,
  in: \emph{Gamma--Ray Bursts: 30 Years of Discovery},
  edited by E.E.\ Fenimore and M.\ Galassi,
  AIP Conf. Proc.  {727}, pp. 684--687 [arXiv:astro-ph/0407144].



\bibitem[Biller et al.(1999)]{Biller}
Biller, S.D., {\it et al.},
1999,
Phys. Rev. Lett.  {83}, 2108 [arXiv:gr-qc/9810044].


\bibitem[Bird et al.(1995)]{Bird-etal1995}
  Bird, D.J., {\it et al.},
  1995,
  Astrophys. J.  {441}, 144 [arXiv:astro-ph/9410067].


\bibitem[Coleman and Glashow(1997)]{ColemanGlashow1997}
Coleman, S.R., Glashow, S.L.,
1997,
Phys. Lett. B {405}, 249 [arXiv:hep-ph/9703240].

\bibitem[Coleman and Glashow(1999)]{ColemanGlashow1999}
Coleman, S., Glashow, S.L.,
1999,
  Phys. Rev. D {59}, 116008 [arXiv:hep-ph/9812418].

\bibitem[Corichi and Sudarsky(2005)]{CorichiSudarsky}
Corichi, A., Sudarsky, D.,
2005,
Int. J. Mod. Phys.  D {14}, 1685
[arXiv:gr-qc/0503078].


\bibitem[Dowker et al.(2004)]{D04}
Dowker, F.,  Henson, J.,  Sorkin, R.D.,
2004,
Mod. Phys. Lett. A  19, 1829  [arXiv:gr-qc/0311055].



\bibitem[Gagnon and Moore(2004)]{GagnonMoore}
Gagnon, O., Moore, G.D.,
2004,
Phys. Rev. D {70},  065002 [arXiv:hep-ph/0404196].


\bibitem[Hawking(1978)]{H78}
Hawking, S.W.,
1978,
Nucl. Phys. B {144}, 349.

\bibitem[Horowitz(1991)]{H91}
Horowitz, G.T.,
1991,
Class. Quant. Grav.  {8}, 587.

\bibitem[Hu(2005)]{Hu05}
Hu, B.L., 2005,
Int. J. Theor. Phys.  {44}, 1785 [arXiv:gr-qc/0503067].


\bibitem[Jacobson et al.(2006)]{JLM:constraints}
Jacobson, T., Liberati, S., Mattingly, D.,
2006,
Ann. Phys. (N.Y.)  {321}, 150  [arXiv:astro-ph/0505267].


\bibitem[Klinkhamer(2005)]{K05}
Klinkhamer, F.R.,
2005,
in: \emph{CP Violation and the Flavour Puzzle:
Symposium in Honour of Gustavo C. Branco},
edited by
\newpage\noindent
D. Emmanuel-Costa \emph{et al.},
Krak\'{o}w, Poligrafia Inspektoratu, pp. 157--191
[arXiv:hep-ph/0511030].

\bibitem[Klinkhamer and Risse(2008)]{KlinkhamerRisse}
Klinkhamer, F.R., Risse, M., 2008,
Phys. Rev.  D  {77}, 117901 [arXiv:0806.4351].

\bibitem[Klinkhamer and Rupp(2004)]{KlinkhamerRuppPRD}
Klinkhamer, F.R., Rupp, C.,
2004,
Phys. Rev. D {70}, 045020 [arXiv:hep-th/0312032].


\bibitem[Klinkhamer and Rupp(2005)]{KlinkhamerRuppPRD-BR}
Klinkhamer, F.R., Rupp, C.,
2005,
Phys. Rev. D {72}, 017901 [arXiv:hep-ph/0506071].


\bibitem[Klinkhamer and Schreck(2008)]{KlinkhamerSchreck}
Klinkhamer, F.R., Schreck, M.,
2008,
Phys. Rev.  D {78}, 085026 [arXiv:0809.3217].

\bibitem[Klinkhamer and Volovik(2005)]{KlinkhamerVolovik}
Klinkhamer, F.R., Volovik, G.E.,
2005,
JETP Lett. {81}, 551 [arXiv:hep-ph/0505033].

\bibitem[Lehnert(2003)]{Lehnert2003}
Lehnert, R.,
2003,
Phys. Rev. D {68}, 085003 [arXiv:gr-qc/0304013].


\bibitem[McEnery et al.(2004)]{McEnery-etal2004}
McEnery, J.E., Moskalenko, I.V., Ormes, J.F.,
2004,
in: \emph{Cosmic Gamma--Ray Sources},
edited by  K.S. Cheng {\it et al.},
(Kluwer Academic, Dordrecht),  pp. 361--395
[arXiv:astro-ph/0406250].

\bibitem[Risse et al.(2004)]{Risse-etal2004}
  Risse, M., {\it et al.},
  2004,
  Astropart. Phys.  {21}, 479 [arXiv:astro-ph/0401629].

\bibitem[Takeda et al.(1998)]{Takeda-etal1998}
  Takeda, M., {\it et al.},
  1998,
  Phys. Rev. Lett.  {81}, 1163 [arXiv:astro-ph/9807193].


\bibitem[Visser(1996)]{V96}
Visser, M.,
1996,
\emph{Lorentzian Wormholes: From Einstein to Hawking}
(Springer, New York). 

\bibitem[Wheeler(1957)]{W57}
Wheeler, J.A.,
1957,
Ann. Phys. (N.Y.)  {2}, 604.

\bibitem[Wheeler(1968)]{W68}
Wheeler, J.A.,
1968,
in: \emph{Battelle Rencontres 1967}, edited by C.M. DeWitt and J.A. Wheeler
(Benjamin, New York), pp. 242--307.

\end{thebibliography}
\end{document}